\documentstyle[emulateapj,epsf]{article}

\slugcomment{Accepted for publication in The Astrophysical Journal Letters}

\begin{document}

\title{Detection of dark matter concentrations in the field of Cl
1604+4304 from weak lensing analysis}
\author{Keiichi Umetsu and Toshifumi Futamase}
\affil{Astronomical Institute, Tohoku University, Sendai 980-8578, 
Japan \\ keiichi@astr.tohoku.ac.jp, tof@astr.tohoku.ac.jp}


\begin{abstract}
We present a weak-lensing analysis of a region around
the 
galaxy cluster Cl 1604+4304 ($z=0.897$) 
on the basis of the deep observations 
with the {\it Hubble Space Telescope} ({\sl HST})/Wide Field Planetary
 Camera 2 (WFPC2).
%
We apply a variant of Schneider's aperture mass technique 
to the observed WFPC2 field
and obtain the distribution of weak-lensing signal-to-noise ratio (S/N) 
within the field.
The resulting S/N map reveals a clear pronounced peak 
located about $1\farcm 7$ ($850h_{50}^{-1}$ kpc at $z=0.897$) southwest of the 
second peak associated with the optical cluster center determined 
from the dynamical analysis of Postman et al. 
A non-linear finite-field inversion method has been used to reconstruct the
projected mass distribution from the observed shear field.
The reconstructed mass map shows a super-critical feature 
at the location of the S/N peak
as well as in the cluster central region. 
Assuming the redshift distribution of field galaxies,
we obtain the total mass in the observed field to be 
$1.0 \times 10^{15} h_{50}^{-1}M_{\odot}$ for 
$\left< z \right>=1.0$.
The estimated mass within a circular aperture of radius $280h_{50}^{-1}$ 
kpc centered on the dark clump
is  $2.4\times 10^{14}h_{50}^{-1}M_{\odot}$. 
We have confirmed the existence of the `dark' mass concentration 
from  another deep {\sl HST} observation with a slightly different
($\sim 20\arcsec$) pointing. 

\end{abstract} 
\keywords{cosmology: observations --- 
dark matter --- 
galaxies: clusters: individual (Cl 1604+4304) --- 
gravitational lensing}


\section{INTRODUCTION}

Weak shear fields of high-redshift galaxies 
are promising, efficient tools
to investigate 
the mass distribution on cluster-supercluster scales   
(Kaiser \& Squires 1993; Luppino \& Kaiser 1997; Kaiser et al. 1998; 
Bartelmann \& Schneider 2000)
and provide a unique mean to 
detect dark mass concentrations (Schneider 1996; Erben et al. 2000).
Wide-field weak-lensing surveys of 
projected mass overdensities can probe the 
statistical clustering properties and underlying cosmology (Bahcall et
al. 1999).
%

Recent observations have revealed the existence of a supercluster at a
high redshift of $z\approx 0.9$ (Lubin et al. 2000). 
This supercluster 
contains two massive galaxy clusters, Cl 1604+4304 at $z=0.897$  and 
Cl 1604+4321 at $z=0.924$. The two clusters are separated by $17\arcmin$
on the sky, corresponding to a projected separation of $\sim
9h_{50}^{-1}$ Mpc. Cl1604+4304 is located at
 ($\alpha_{{\rm J}2000}$, $\delta_{{\rm J}2000}$) 
 = ( $16{\rm h}\,04{\rm m}\,19.5{\rm s}$,
    $+43\arcdeg\,04\arcmin\,33\farcs 9$) and 
one of the optically-selected high-redshift cluster candidates  
studied by Oke, Postman, \& Lubin (1998). 
Postman, Lubin, \& Oke(1998) presented a detailed 
photometric and spectroscopic survey of the cluster and obtained 
a velocity dispersion of $1226$ km s$^{-1}$ and a dynamical 
mass estimate 
of $6.2\times 10^{15}h_{50}^{-1}M_{\odot}$.  

In this Letter, we present a weak lensing analysis of the 
Cl 1604+4304 field on the basis of deep images taken with the
{\it Hubble Space Telescope} ({\sl HST})/Wide Field Planetary Camera 2
(WFPC2).  
We shall mainly describe the weak-lensing signal-to-noise ratio (S/N)
analysis of the observed field; the details of the
mass reconstruction procedure will
be present elsewhere.
Throughout this Letter, we adopt $\Omega_0=1$ ,$\Omega_{\Lambda}=0$, and 
$H_0=50h_{50}$ km s$^{-1}$ Mpc$^{-1}$; 
$1\arcmin$ on the sky corresponds to $0.52h_{50}^{-1}$ Mpc at the
cluster redshift.
\section{OBSERVATIONS AND DATA REDUCTION}

%
%

The cluster Cl 1604+4304 field was observed in 1994 and 1995
using the WFPC2 camera with the F814W filter on board 
the {\sl HST} (P.I.: J. Westphal; proposal 5234). 
We retrieved the calibrated data for Cl 1604+4304 field 
from the {\sl HST} archive.
Both the 1994 and 1995 observations have the same total exposure time
of $32$ ksec, consisting of 16 single orbits of $2$ ksec.  
The 1995 pointing covers the central region of Cl 1604+4304 (Lubin et
al. 1998), while the 1994 pointing is about $20\arcsec$
offset ($\Delta X\simeq 0\farcs 1,\Delta Y\simeq 18\farcs 2$)
from the 1995 pointing. 
For this reason,
Lubin et al. (1998) discarded the 1994 data from their analysis. 
We analyze both the 1994 and 1995 data separately.
For each observation, the data were shifted and combined into 
the final frame to remove cosmic rays 
using the IRAF/STSDAS task CRREJ. 
The PC chip was discarded 
from our analysis because of its brighter isophotal limit, so that
the final frame for each observation consists of three WFC chips.
The side length of the WFPC2 field is about $2\farcm 5$ (
$1.26 h_{50}^{-1}$ Mpc at $z=0.897$).

We use the SExtractor package (Bertin \& Arnouts 1996) 
for the object detection, photometry, and measurement of image shapes.
We extract all objects with isophotal areas
larger than 12 pixels ($0\farcs 0996$ pixel$^{-1}$) 
above $2\sigma$ pixel$^{-1}$ of the local sky level. 
We calculate
the total magnitudes $I_{814{\rm W}}$ in STMAG system for each object
using MAG\_BEST parameters in SExtractor.
The corresponding detection thresholds for the 1994 and 1995 data  are 
$\mu_{\rm 814W}=26.2$ and
$26.5$ mag arcsec$^{-2}$, respectively.
We measure the quadruple moments 
$Q_{ij}=\int\!d^2\theta\, \theta_i\theta_j I(\vec{\theta})/\int\!d^2\theta\, 
I$ of the surface brightness $I(\vec{\theta})$ for each object,
where the center of light is chosen as the coordinate origin. From 
$\{Q_{ij}\}$, we construct the image ellipticity 
$\epsilon_i= ( Q_{11}-Q_{22},2Q_{12} )/
({\rm Tr}(Q)+2({\rm det}Q)^{1/2})$ $(i=1,2)$.
To construct catalogs of faint galaxies,
we refer to the following SExtractor output parameters:
We exclude objects with FLAGS 
(extraction flags) $\ge 1$ from our analysis. 
We make star/galaxy separation 
using the 
CLASS\_STAR parameter (stellarity index)
and keep the objects with ${\rm CLASS\_STAR}\le 0.2$ as faint galaxy
candidates. 
Objects with
FWHM\_IMAGE (FWHM profile from a Gaussian fit to the core) 
$<4$ pixels are excluded
since image shapes of extremely small objects relative to the pixel
scale 
may be affected by the anisotropic PSF.
Finally, 
all objects with  
$I_{814{\rm W}}\in(25.5,27.5)$ are selected as background galaxy
candidates.  
This detection and selection procedure leads to 
the final catalogs 
with total galaxy numbers of
$N_{\rm g}=$ 
200 and 241,
corresponding number densities 
of $n_{\rm g}=$ $45.8$ and $56.8$ arcmin$^{-2}$,
for the 1994 and 1995 data, respectively.

\section{WEAK LENSING ANALYSIS}

\subsection{Strategy}
Our first aim is to obtain the distribution of weak-lensing S/N
in the data field.
Then, the high peaks in the S/N maps can be 
identified as clusters or mass overdensities.
To do this, we make use of a variant of aperture mass statistics.
The statistics rely on the fact that the shear 
$\gamma=(\gamma_1,\gamma_2)$ 
and the convergence
$\kappa:=\Sigma/\Sigma_{\rm cr}$
are 
related to each other through $\vec{\nabla} \kappa
=\hat{D}\gamma
\equiv \vec{u}_{\gamma}$, 
where 
$\Sigma$ is the surface mass density of the deflector,
$\Sigma_{\rm cr}=(c^2/4\pi G)D_{\rm s}/D_{\rm d}D_{\rm ds}$ 
%
%
is the critical surface mass density, and $\hat{D}=\{\hat{D}_{ij}\}$ 
($i,j=1,2$) is a differential operator defined by
\begin{equation}
\hat{D}= 
\left(
\begin{array}{@{\,}cc@{\,}}
 \partial/\partial\theta_1 & \partial/\partial\theta_2 \\
-\partial/\partial\theta_2 & \partial/\partial\theta_1
\end{array}
\right)
\end{equation}
(Kaiser 1995). 
Since $\vec{u}_{\gamma}$ is the gradient of $\kappa$, 
operating $\hat{D}$ further on $\vec{u}_{\gamma}$ yields 
\begin{equation}
\label{eq:divrot}
(\triangle\kappa,0)=
\hat{D}^2\gamma=\hat{D}\vec{u}_{\gamma}=({\rm div}\vec{u}_{\gamma},
{\rm rot}\vec{u}_{\gamma}).
\end{equation}
In weak lensing limit ($\kappa\ll1$ and $|\gamma|\ll 1$), 
the expectation value of the image ellipticity, 
${\rm E}[\epsilon(\vec{\theta})]$,
is the shear $\gamma(\vec{\theta})$.
In practice, however, background sources have intrinsic ellipticities 
$\epsilon_{({\rm s})}$, so that
weak lensing analysis involves the smoothing procedure to reduce 
the noise. 
We denote the smoothed fields by angular brackets $\left< \ \right>$: e.g.,
$\left<\kappa\right>=\int\,d^2\theta'\,
W(|\vec{\theta}-\vec{\theta}'|;\vartheta)\,\kappa(\vec{\theta}')$,
where $W(\theta;\vartheta)$ is a smooth, continuous window function with
a characteristic scale of $\vartheta$.
Because of the commutativity between smoothing and the 
mass reconstruction (Van Waerbeke 2000), 
the smoothed quantities $\left<\kappa\right>$ and 
$\left<\gamma\right>$ satisfies the same relations as those between 
$\kappa$ and $\gamma$:
$\vec{\nabla}\left<\kappa\right>=\hat{D}\left<\gamma\right>
\equiv \vec{u}_{\left<\gamma\right>}$
;
$(\triangle\left<\kappa\right>,0)=
\hat{D}^2\left<\gamma\right>=\hat{D}\vec{u}_{\left<\gamma\right>}
=({\rm div}\vec{u}_{\left<\gamma\right>},
{\rm rot}\vec{u}_{\left<\gamma\right>})$. 
The Laplacian of 
$\left<\kappa\right>$, i.e. ${\rm div}\vec{u}_{\left<\gamma\right>}$,
is just the convergence convolved with 
the compensated filter function $\triangle W$, 
and hence equivalent to the aperture mass.
Defining  $\vec{u}_{\left<\epsilon\right>}
:=\hat{D}\left<\epsilon\right>$, we see that
the observable ${\rm div}\vec{u}_{\left<\epsilon\right>}$
traces the distribution of
$\Sigma$ in the limit of weak lensing,
as pointed out by Luppino \& Kaiser (1997).
On the other hand, ${\rm rot}\vec{u}_{\left<\epsilon\right>}$ measures
the `pure' noise.
%
%
%
%
%
A discretized estimator for $\hat{D}\vec{u}_{\left<\gamma\right>}$ is
given by
$\hat{D}\vec{u}_{\left<\epsilon\right>}=
({\rm div}\vec{u}_{\left<\epsilon\right>},
 {\rm rot}\vec{u}_{\left<\epsilon\right>})=
-n_{\rm g}^{-1}\sum_{m=1}^{N_{\rm
g}}p(|\vec{\theta}-\vec{\theta}_m|;\vartheta)\,
[\epsilon_{\rm t}(\vec{\theta}_m;\vec{\theta}),
 \epsilon_{\rm r}(\vec{\theta}_m;\vec{\theta})] 
$, where $\epsilon_{\rm t}$ and $\epsilon_{\rm r}$ are the tangential and
the radial components of the image ellipticity 
$\epsilon=\epsilon_1+i\epsilon_2$
defined by
$\epsilon_{\rm t}(\vec{\theta};\vec{\theta}_0):=-
\Re[\epsilon(\vec{\theta})e^{-2i\phi}]$ and
$\epsilon_{\rm r}(\vec{\theta};\vec{\theta}_0):=-
\Im[\epsilon(\vec{\theta})e^{-2i\phi}]$ with 
$\phi= {\rm Arg}(\vec{\theta}-\vec{\theta}_0)$ 
and
$p(\theta;\vartheta)=W''(\theta;\vartheta)-W'(\theta;\vartheta)/\theta$.
The noise properties of $\hat{D}^2\left<\epsilon\right>$ 
due to the intrinsic source ellipticities 
are contained in the covariant matrix $\sigma^2_{ij}:={\rm E}[
\hat{D}^2\left<\epsilon^{\rm (s)}\right>\,
\hat{D}^2\left<\epsilon^{\rm (s)}\right>]_{ij}$:
\begin{equation}
\label{eq:sigma}
\sigma^2_{ij}(\vec{\theta})=\delta_{ij}
\frac{\sigma_{\epsilon}^2}{2n_{\rm g}^2}
\sum_{m=1}^{N_{\rm g}} 
p^2
(|\vec{\theta}-\vec{\theta_m}|;\vartheta)\equiv \delta_{ij}
\sigma^2(\vec{\theta}), 
\end{equation}
where $\sigma_{\epsilon}$ is the dispersion of the intrinsic source
ellipticities and
we have assumed ${\rm E}[\epsilon^{\rm (s)}(\vec{\theta}_m)
\epsilon^{\rm (s)}(\vec{\theta}_n)]_{ij}=\delta_{ij}\delta_{mn}
\sigma_{\epsilon}^2/2$.
The Kronecker delta $\delta_{ij}$ in equation (\ref{eq:sigma}) 
ensures
that the dispersions of 
${\rm div}\vec{u}_{\left<\epsilon\right>}$  
and
${\rm rot}\vec{u}_{\left<\epsilon\right>}$ 
are the same and that the two fields are statistically uncorrelated.
The local weak-lensing S/N  
$\nu(\vec{\theta})$ at position $\vec{\theta}$ is 
then defined by
\begin{equation}
\label{eq:S/N}
\nu(\vec{\theta})
:=-\frac{{\rm div}\vec{u}_{\left<\epsilon\right>}
(\vec{\theta})}{\sigma(\vec{\theta})}
=\frac{\sqrt{2}}{\sigma_{\epsilon}}\frac{\sum_m 
p(|\vec{\theta}-\vec{\theta}_m|;\vartheta)
\epsilon_{\rm t}(\vec{\theta}_m;\vec{\theta})}
{\sqrt{\sum_m p^2(|\vec{\theta}-\vec{\theta_m}|;\vartheta)}}.
\end{equation} 
The resulting formula (eq. [\ref{eq:S/N}])
is equivalent to the one derived
by Schneider (1996) using aperture mass statistics. 
In the present Letter, we use a Gaussian window function of the form 
$W_{\rm
G}(\theta;\vartheta)=\exp(-\theta^2/\vartheta^2)/\pi\vartheta^{2}$,
in which case $p(\theta;\vartheta)=4(\theta/\vartheta)^2
\exp(-\theta^2/\vartheta^2)/\pi\vartheta^4\equiv p_{\rm
G}(\theta;\vartheta)$; $p_{\rm G}(\theta;\vartheta)$ has its maximum at
$\theta =\vartheta$ and falls off rapidly at $\theta>\vartheta$. 
These statistics can be applied to the strong-lensing
regime (e.g., cluster central region),
because 
the contribution of image ellipticities
to the aperture mass comes mainly from galaxies
within an annulus at radius $\vartheta$ and thus we can avoid the strong
lensing regime.
For each independent {\sl HST} observation,
we perform a local S/N analysis using equation (\ref{eq:S/N}). 

To obtain the projected mass distribution,
we perform a mass reconstruction to the {\sl HST}/WFC field.
Taking into account the high redshift ($z=0.897$) of 
Cl 1604+4304 and 
small field-of-view ($2\farcm 5$ on a side) of the {\sl HST}/WFC field,
we adopt a non-linear finite-field inversion method developed by 
Seitz \& Schneider (1997),
which takes account of the source redshift distribution.
Since little is known about
the redshift distribution of field galaxies,
we assume a source redshift distribution 
of the form 
$p_{z}(z)=\beta z^2 \exp[-(z/z_0)^{\beta}]/\Gamma(3/\beta)z_0^3$
(Brainerd, Blandford, \& Smaili 1996), in which case
the mean redshift $\left<z\right>$ is given by 
$\left<z\right>=z_0\Gamma(4/\beta)/\Gamma(3/\beta)$. 
In the present Letter, we consider only the case
$(\left<z\right>, \beta)=(1.0, 1.0)$.

Once a smoothed ellipticity field $\left<\epsilon\right>(\vec{\theta})$
is obtained from the observed image ellipticities,
a convergence map can be obtained through the integral equation 
\begin{equation}
\label{eq:MR}
\kappa_{\infty}(\vec{\theta})-\bar{\kappa}_{\infty}
=\int_{\cal U}\!d^2\theta'\,
\vec{H}(\vec{\theta},\vec{\theta}')\cdot\vec{u}_{\gamma_{\infty}}
(\vec{\theta}'),
\end{equation}
where quantities with $\infty$-subscript represent the values for sources 
at infinite redshift,
$\bar{\kappa}_{\infty}$ is the unknown constant 
which represents the average of
$\kappa_{\infty}$  within the data field $\cal U$,
$\vec{H}$ is the kernel which is
the gradient of the scalar field that satisfies the
Neumann boundary problem (see Seitz \& Schneider 1996), and
$\vec{u}_{\gamma_{\infty}}=\hat{D}\gamma_{\infty}$. 
In general, the shear is not direct observable, so that 
the integral equation (\ref{eq:MR}) is non-linear and solved
iteratively: A mass reconstruction scheme applied to the irregular {\sl
HST}/WFC field is outlined in Seitz et al. (1996) and Umetsu, Tada, \&
Futamase (1999).
%

In order to determine the constant $\bar{\kappa}_{\infty}$, 
we employ a dynamical mass estimate of CL 1604+4304 obtained by
Postman et al. (1998).  Postman et al. (1998) estimated the projected
mass inside the circular aperture of radius
$R_{500}=500h_{50}^{-1}$ kpc 
at the dynamical center
to be $M_{500}=0.842\times 10^{15}h_{50}^{-1}M_{\odot}$,
corresponding to the
mean convergence within the circular aperture of 
$\bar{\kappa}_{\infty,500}=M_{500}/\pi R_{500}^2\Sigma_{\rm cr,\infty}=0.811$.
However, owing to the small field-of-view of the {\sl HST}/WFPC2, 
the observed data fields do not cover the whole aperture area 
even for the 1995 pointing. 
The effective aperture area covered by the 1995 data field
is about $60\%$ of the circular aperture area. 
We assume that the mean
convergence $\bar{\kappa}_{\infty,{\rm eff}}$ 
inside the effective aperture 
is the same as the mean convergence $\bar{\kappa}_{\infty,500}$
inside the circular aperture: 
$\bar{\kappa}_{\infty,{\rm eff}}\approx\bar{\kappa}_{\infty,500}=0.811$.
We, thus, solve the integral equation (\ref{eq:MR}) iteratively 
varying the value of $\bar{\kappa}_{\infty}$
in such a way  that
the average $\bar{\kappa}_{\infty,{\rm eff}}$ of the
reconstructed convergence within the effective aperture area equals 
$\bar{\kappa}_{\infty,500}=0.811$. 

\subsection{Analysis of Cl 1604+4304}

Since the 1995 pointing covers the central cluster region
where a dynamical analysis was performed by Postman et al. (1998),
we first analyzed the 1995 data. 
We calculated the fields $\hat{D}\vec{u}_{\left<\epsilon\right>}$
%
with $\vartheta=0\farcm 4$ ($150h_{50}^{-1}$ kpc)
on $24\times 24$ grid points $\vec{\theta}_{ij}$
(except on the PC chip), 
from which  
we obtained the distribution of
the weak-lensing S/N within the WFC field.
Here we have measured the dispersion $\sigma_{\epsilon}$ 
from the rotation field 
${\rm rot}\vec{u}_{\left<\epsilon\right>}(\vec{\theta}_{ij})$ in the
following way: 
We calculated the variance
$\sigma^2_{\rm rot}$ of the noise field ${\rm
rot}\vec{u}_{\left<\epsilon\right>}(\vec{\theta}_{ij})$ 
by averaging 
over the grid points.
Equating $\sigma^2_{\rm rot}$ with 
the average of $\sigma^2(\vec{\theta})$ 
over the grid points $\vec{\theta}_{ij}$,
we obtained an estimate of $\sigma_{\epsilon}=0.21$. 

Figure 1
displays the resulting 
weak-lensing S/N map obtained for the 1995 data. 
We find two S/N peaks above a threshold of $\nu_{\rm th}=3$. 
The first maximum has a peak height of $\nu=4.5$ and its location is
about $1\farcm 7$ ($850h_{50}^{-1}$ kpc) southwest of the
the optical cluster center (marked with cross)
determined from
the dynamical analysis of Postman et al. (1998).
%
%
%
The peak hight of the second maximum is $\nu=3.6$ and 
located in the vicinity of the dynamical cluster center.
This offset is about $0\farcm 15$ ($75h_{50}^{-1}$ kpc) and 
comparable to the grid spacing of $0\farcm 1$ ($50h_{50}^{-1}$ kpc),
so that the second peak is 
consistent with the dynamical cluster center
and associated with the cluster member galaxies (Lubin et al. 1998).
In this field, there are a total of 12 cluster members whose redshifts are
spectroscopically confirmed by Keck observations
(Lubin et al. 1998). 
However, 
no cluster members are observed at around the first S/N peak 
(see Fig. 4 of Lubin et al. 1998) because 
objects located on
the upper edge of the WF4 chip (upper-right WF chip)
are not contained within the
sampled region of spectroscopic observations (Postman et al. 1998).

Although the detection of the mass concentration cannot be confirmed
spectroscopically from the available data,
we have checked the stability and reliability 
of S/N maps.
We find that the main features, i.e. the locations and heights of the peaks,  
are insensitive to the choice of the weight function 
or filtering scale. 
It shoud be noted that these main features are stable
even if objects with large ellipticities ($|\epsilon|\ge 0.4$) 
are removed from our galaxy catalog, in which case
the peak heights of the first and second
maxima are $3.7$ and $3.0$, respectively. 
There still remains a possibility that the detected S/N peaks might 
be noise peaks due to the intrinsic source ellipticities or the anisotropic 
PSF. Van Waerbeke (2000) has shown from the simulated data
that the probability distribution
of the noise peaks in mass maps deviates from Gaussian and 
the probability of finding high noise peaks increases owing to the
boundary effects. 
Since the 
field-of-view of the {\sl HST}/WFC is small 
and we adopt a constant filtering scale, the obtained
S/N maps may be affected by boundary effects.
We can investigate such possibilities using independent
deep observation of this field.
In Fig. 2,  
we show the weak-lensing S/N map obtained from the 1994 data.
Here we have used the same filtering scale 
and grid spacing 
as for the 1995 data. 
Comparing Fig. 2 with Fig. 1, we see that the first maximum in the
1994 S/N map corresponds to that in the 1995 S/N map. 
The first maximum in the 1994 S/N map has a peak height of $\nu=4.2$,
whose location 
coincides well with that of the first maximum in the 1995 S/N
map
($\alpha_{\rm J2000}=16{\rm h}\,04{\rm m}\,15.7{\rm s}$ and 
$\delta_{\rm J2000}=+43\arcdeg\,03\arcmin\,47\farcs 2$).
The identification of the first S/N maximum 
from two-different pointing observations 
indicates that the S/N maximum does not originate from the
anisotropic PSF. 
Moreover, since the first maximum in the 1994 S/N map
is located at the center of the WF4 chip, 
the S/N measure at the first peak cannot be affected by boundary effects. 
These results confirm the detection of `dark' 
mass concentration associated with the first S/N maximum. 

We have reconstructed the projected mass distribution within the 1995
data field. 
We calculated the smoothed ellipticity 
$\left<\epsilon\right>(\vec{\theta})$  on $20\times 20$ grid points 
(except on the PC chip) using $W_{\rm G}(\theta;\vartheta)$ with 
$\vartheta = 0\farcm 35$ ($176h_{50}^{-1}$ kpc). 
In Fig. 3, we show the reconstructed $\kappa_{\infty}$-map. 
In mass reconstruction, we implicitly assumed that
the dark mass concentration has the same redshift $z=0.897$ as 
Cl 1604+4304.
We see from Figs. 1 and 3 
that the local maxima in the S/N maps correspond
to the mass peaks and that the weak-lensing S/N traces the projected
mass, as expected.  The $\kappa_{\infty}$-map shows a super-critical
feature around the dark mass concentration ($\kappa_{\infty}=1.19$)
as well as in the cluster central region.
The projected mass within the 1995 data field is obtained as
$1.0 \times
10^{15}h_{50}^{-1}M_{\odot}$. From the mass map,
the enclosed mass in a $280h_{50}^{-1}$ kpc radius
centered on the dark mass concentration is calculated to be
$2.4\times 10^{14}h_{50}^{-1}M_{\odot}$: 
For the mass-to-light ratio,
a lower limit is obtained as $M/L_{\rm B}\approx 500h_{50}$ in solar
units from an upperlimit on the dark clump's surface brightness.

\section{DISCUSSION AND CONCLUSIONS}

   
We have performed a weak-lensing analysis
on the deep images of
a region around Cl 1604+4304 taken with the {\sl HST}/WFPC2.
We detected two significant maxima in the resulting weak-lensing S/N
map: 
the second peak associated with the dynamical cluster center and
the first peak located about $1\farcm 7$ south of the dynamical
center having no optical counter parts.  
The identification of high S/N peak from two-different pointing
observations provides a robust evidence of
dark mass concentration. 
Assuming that the dark mass clump has the same redshift
as Cl 1604+4304, 
we reconstructed the projected mass distribution in the
1995 data field, from which we have estimated the
mass within a circular aperture of radius 
$280h_{50}^{-1}$ kpc centered on the dark clump to be
$2.4\times 10^{14}h_{50}^{-1}M_{\odot}$. 
The obtained results  
indicate that this dark
clump could be a compact, massive bound system 
associated with the supercluster at $z\approx 0.9$; 
the obtained mass will be overestimate if the redshift of the dark
clump is much less than $0.897$.

Our results demonstrate that the weak-lensing
S/N statistics are powerful and efficient tools even for 
high-redshift cluster surveys ($z\sim 1$).
For a further confirmation of the dark mass concentration, 
deep wide-field observations covering the entire supercluster field
($\sim 20\arcmin$) are required. Such observations can probe the mass 
distribution in the supercluster field extending over $\sim
10h_{50}^{-1}$ Mpc. 


\acknowledgements

We are very grateful to Toru Yamada, Masahiro Takada, and Masaru Kajisawa 
for helpful discussions. We thank the anonymous referee for useful
comments. 
Part of this work is based on the observation with the NASA/ESA {\it Hubble}
{\it space} {\it Telescope}, obtained from the data archive at the Space 
Telescope Science Institute (STScI). STScI is operated by the
Association of Universities for Research in Astronomy, Inc. under the
NASA contract NAS 5-26555. 



\begin{figure}
\epsfxsize=160mm
\epsfbox{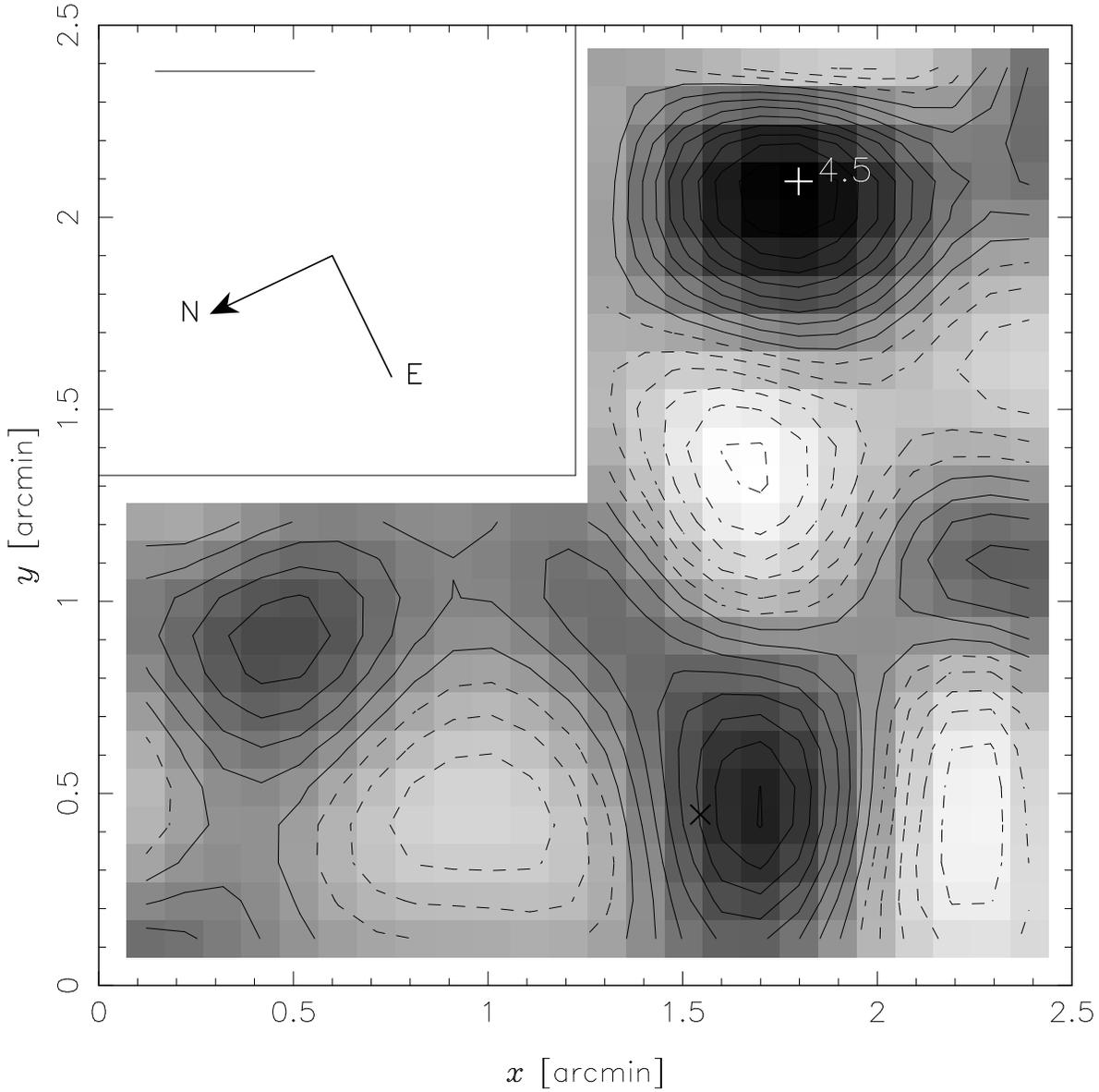}
\figcaption{Contour plot of the weak-lensing signal-to-noise ratio
distribution 
obtained from 241 galaxy images with $I_{814{\rm W}}\in (25.5,27.5)$
taken with the {\sl HST}/WFPC2 in 1995. 
The solid and dashed lines indicate the positive and negative ones,
respectively. The contours are stepped in units of $0.5$. 
The filtering scale ($0\farcm 4$) used for the
aperture mass measure is marked and
the position of the dynamical cluster center is indicated ($\times$). 
The location of the first maximum is marked with $+$.
The peak height of the second maximum is $3.6$ and
its location coincides well with the dynamical center. 
 } 
\end{figure}

\begin{figure}
\epsfxsize=160mm
\epsfbox{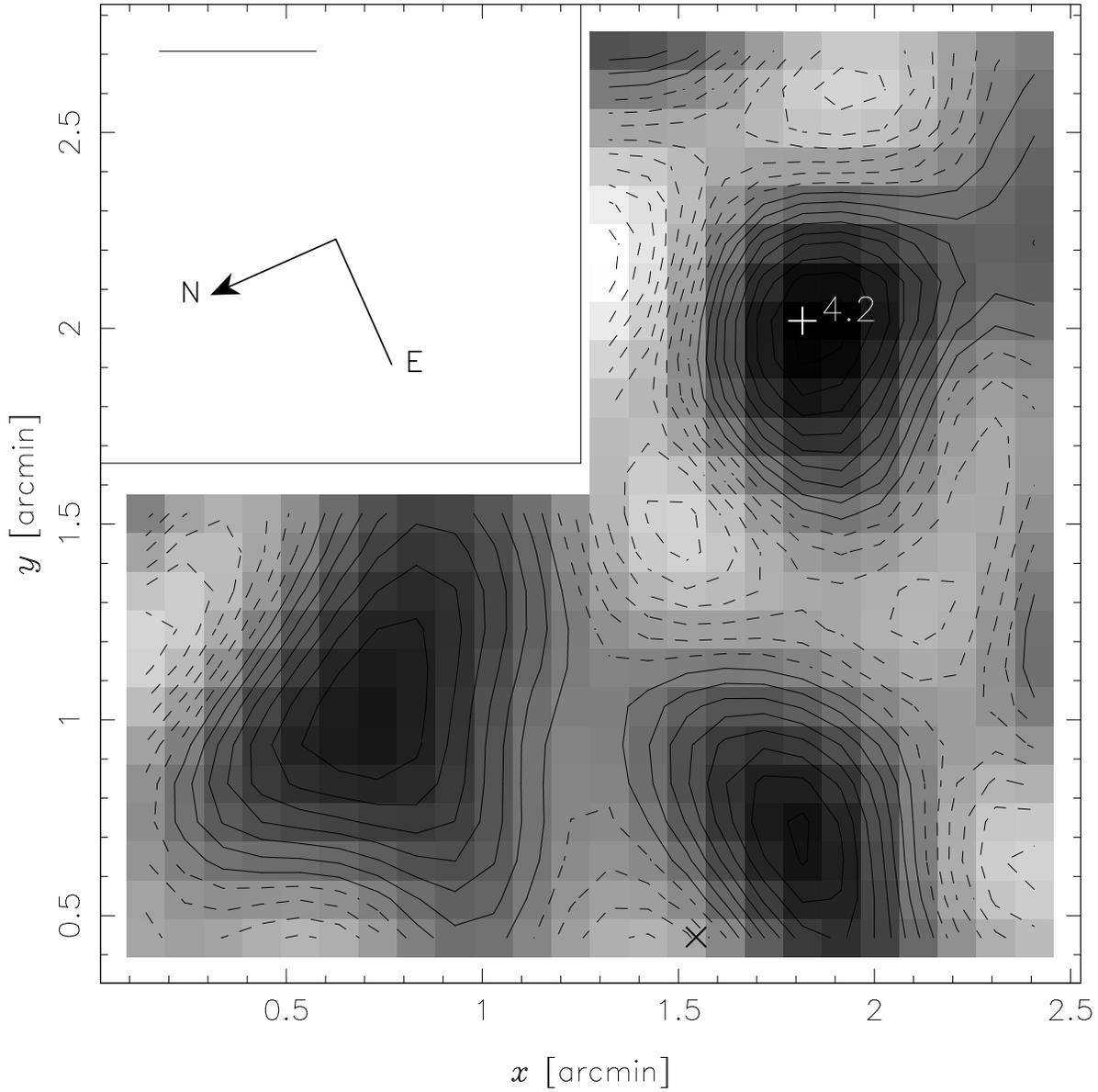}
\figcaption{The same as Fig. 1 but 
obtained from 200 galaxy images with $I_{814{\rm W}}\in (25.5,27.5)$
taken with the {\sl HST}/WFPC2 in 1994. 
The same  coordinate system as in Fig. 1 is used.
} 
\end{figure}

\begin{figure}
\epsfxsize=160mm
\epsfbox{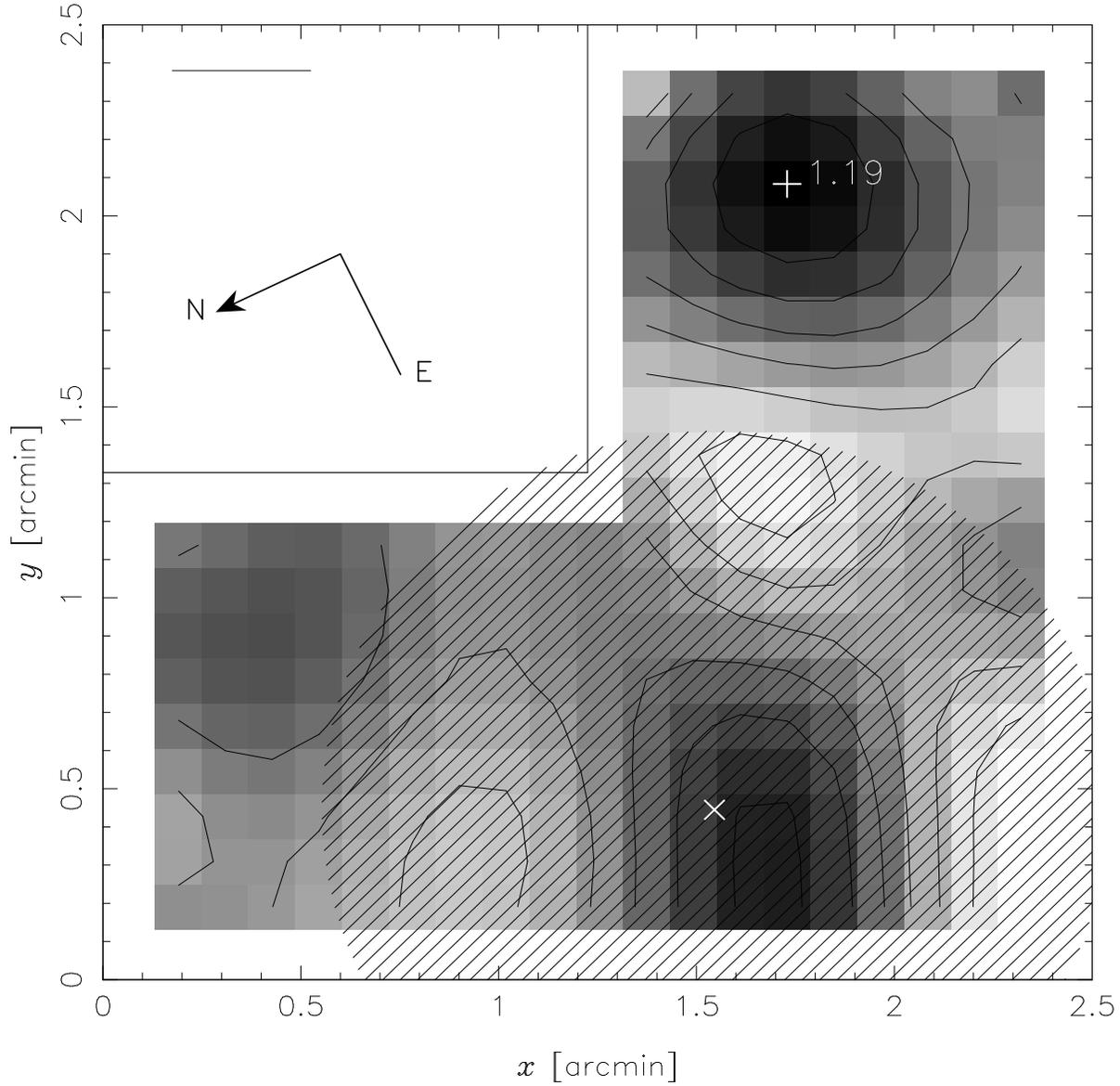}
\figcaption{ Contour plot of the 
$\kappa_{\infty}$-distribution reconstructed from 
the shear field of the 1995 observation.
The contours are stepped in units of $0.1$. 
The smoothing scale ($0\farcm 35$) used to calculate the
shear field    
is marked. The shaded area indicates the circular aperture of radius
$500h_{50}^{-1}$ kpc centered on the dynamical cluster center (marked
with $\times$).
The side length is about $2\farcm 5$ ($1.26h_{50}^{-1}$ 
Mpc at $z=0.897$).
The mass distribution is super-critical around the first maximum (marked 
with $+$) and the cluster central region.
} 
\end{figure}


\begin{thebibliography}{99}

\bibitem{BAHCALL}
Bahcall, N. A., Ostriker, J. P., Perlmutter, S., Steinhardt,  P. J. 1999,
Sci 284, 1481B

\bibitem{BART}
Bartelmann, M. \& Schneider, P. 2000, \physrep, submitted

\bibitem{BA96}
Bertin, E. \& Arnouts, S. 1996, A\&AS 117, 393 

\bibitem{Br96}
Brainerd, T. G., Blandford, R. D.,  \&  Smail, I. 1996
ApJ, 466, 623

\bibitem{Erben} Erben, T., Van Waerbeke, L., Mellier, Y., Schneider, P., 
	Cuillandre, J.-C., Castander, F. J., \& Dantel-Fort, M. 2000,
	A\&A, 355, 23 


\bibitem{K95} Kaiser 1995, ApJ, 439, L1

\bibitem{KS} Kaiser, N. \& Squires, G. 1993, ApJ, 404, 441

\bibitem{K1998} Kaiser, N., Wilson, G., Luppino, G., Kofman, L., 
Gioia, I., Metzger, M., \& Dahle, H. 1998, preprint astro-ph/9809268


\bibitem{LUBIN98} Lubin, L., Postman, M., Oke, J. B., Ratnatunga, K. U., 
	Gunn, J. E., Hoessel, J. G., \& Schneider D. P. 1998, ApJ, 116, 584 

\bibitem{LUBIN00} Lubin, L., Brunner, R., Metzger, M. R.,
Postman, M., Oke, J. B, 2000, ApJ, 531, L5 

\bibitem{LK97} Luppino, G. A. \& Kaiser, N. 1997, ApJ, 475, 20

\bibitem{OKE98} Oke, J. B., Postman, M., \& Lubin, L. 1998, ApJ, 116, 549 

\bibitem{POST98} Postman, M., Lubin, L., \& Oke, J. B. 1998, ApJ, 116, 560 


\bibitem{S96} Schneider, P. 1996, MNRAS, 283, 837

\bibitem{SS96} Seitz, C. Kneib, J.-P., Schneider, P., \& Seitz, S.
 1996, A\&A, 314, 707

\bibitem{SS95} Seitz, S. \& Schneider, P. 1996, A\&A, 305, 383

\bibitem{SS97} Seitz, C. \& Schneider, P. 1997, A\&A, 318, 687

\bibitem{UTF} Umetsu, K., Tada, M., \& Futamase, T. 1999,
	Prog. Theor. Phys. Suppl., 133, 53

\bibitem{VW} Van Waerbeke, L. 2000, MNRAS, 313, 524
\end{thebibliography}
\end{document}